\documentclass[pre,aps,twocolumn,showpacs]{revtex4-2}
\usepackage{graphicx,epstopdf}
\usepackage{dcolumn}
\usepackage{bm}
\usepackage{color}
\usepackage{xcolor}
\usepackage{float}
\usepackage{amssymb,amsmath}
\begin{document}

\title{Universality in conformations and transverse fluctuations of a semi-flexible polymer in a crowded environment}

\author{Jacob Bair}
\author{Swarnadeep Seth}
\author{Aniket Bhattacharya}
\altaffiliation[]
{Author to whom the correspondence should be addressed}
{}
\email{Aniket.Bhattacharya@ucf.edu}
\affiliation{Department of Physics, University of Central Florida, Orlando, Florida 32816-2385, USA}
\date{\today}
\begin{abstract}
 We study universal aspects of polymer conformations and transverse
 fluctuations for a single swollen chain characterized by a contour
 length $L$ and a persistence length $\ell_p$ in two
 dimensions (2D) and  in three dimensions (3D) in the bulk, as well as in 
 the presence of excluded volume (EV) particles of different sizes occupying
 different volume fractions. In the absence of the EV particles we
 extend the previously established universal scaling relations in 2D  [A. Huang, A. Bhattacharya, and K. Binder,
J. Chem. 140, 214902 (2014)] to include 3D and demonstrate that the scaled
 end-to-end distance $\langle R_N^2\rangle/(2 L\ell_p)$ and
 the scaled transverse fluctuation $\sqrt{\langle{l_{\perp}^2}\rangle}/{L}$ as a
 function of $L/\ell_p$ collapse onto the same master curve, where  $\langle R_N^2\rangle$ and
 $\langle{l_{\perp}^2\rangle}$ are the
 mean-square end-to-end distance and 
transverse fluctuations. 
However, unlike in 2D, where the Gaussian regime is absent due to
extreme dominance of the EV interaction, we find the Gaussian regime
is present, albeit very narrow in 3D. The scaled transverse
fluctuation in the limit $L/\ell_p \ll 1$ is independent of the
physical dimension and scales as  $\sqrt{\langle{l_{\perp}^2}\rangle}/{L} \sim
(L/\ell_p)^{\zeta-1}$, where $\zeta = 1.5$ is the roughening
exponent. For $L/\ell_p \gg 1$ the scaled fluctuation scales as
$\sqrt{\langle{l_{\perp}^2}\rangle}/{L} \sim
(L/\ell_p)^{\nu-1}$, where $\nu$ is Flory
exponent for the corresponding spatial dimension ($\nu_{2D}=0.75$, and
$\nu_{3D}=0.58$). 
When EV particles of different sizes at different densities are added
into the system, our results indicate that the crowding density either
does not or only weakly affects the universal scaling relations. We
discuss the implications of these results in living matter by showing the
experimental result for a dsDNA on the master plot.
\end{abstract}
\maketitle
\section{Introduction}
Polymers in a crowded environment is a ubiquitous process in living matter~\cite{Review1,Review2,Review3}. Molecular crowding significantly affects
the structure and function of bio-macromolecules. For example, individual DNA molecules in a crowded environment have been observed to undergo compactification in the presence of negatively charged
proteins~\cite{DNA_collapse}. Likewise, in the industrial world the presence
of nanoparticles affects the static phase diagram and dynamics of the
polymer-nano-composites in a nontrivial way. Various factors, {\em
  e.g.}, the volume fraction of the crowding species, the strength of
the polymer-particles, intra-polymer and intra-particle interactions,
the temperature, the contour length $L$, and the persistence length
$\ell_p$ affect its statics and dynamics. Thus it is useful to use scaling theories
of polymers~\cite{deGennes} to understand the inter dependencies of
various factors towards a universal theory of polymer conformations
and dynamics in such systems. This approach also helps to plot 
experimental data in terms of scaled quantities to develop a better 
understanding of the experimental system
studied~\cite{Moukhtar_PRL_2007}-\cite{Microtubules}. \par
In order to understand experimental data, biopolymers are typically
described by a Worm-Like-Chain (WLC) Kratky-Porod
model~\cite{Rubinstein,Kratky} whose 
 mean square end-to-end distance $\langle R_N^2\rangle$ is given by~\cite{Rubinstein}
\begin{equation}
\frac{\langle R_N^2\rangle}{L^2} = \frac{2\ell_p}{L}\left(1-\frac{\ell_p}{L}[1-\exp(-L/\ell_p)]\right).
\label{WLC}
\end{equation}
For  $\ell_P \gg L$, $\langle R_N^2\rangle = L^2$ and the chain behaves
like a rod, while  
for $L \gg \ell_p $ the limiting behavior of the WLC is that of a Gaussian chain ($\langle R_N^2\rangle = 2L\ell_p$). 
However, it is expected that for $L \gg \ell_p $ the chain will eventually feel the effect of the EV interaction and will exhibit the conformation statistics for a 
swollen chain that is not captured in the WLC description.  
Indeed, we know from theoretical arguments following Schaefer {\em
  et al.}~\cite{Pincus_MM_1980} and Nakanishi~\cite{Nakanishi_1987}
that a proper
description of an EV swollen chain in $d$ spatial dimensions is given by 
\begin{equation}
\sqrt {\langle R_N^2 \rangle} \simeq b_l^{\frac{d-2}{d+2}} N^{\frac{3}{d+2}}\ell_p^{\frac{1}{d+2}} = b_l^{\frac{d+1}{d+2}}\left( \frac{L}{b_l}\right)^{\nu}\ell_p^{\frac{1}{d+2}}. 
\label{Rn_EV}
\end{equation}
Here $N$ is the number of monomers of the chain so that $L = (N-1)b_l
\simeq Nb_l$ (for $N \gg 1 $), $b_l$ is the bond length between two
neighboring monomers, and the mean field Flory exponent $\nu =
3/(d+2)$ in 2D = 0.75 and in 3D = 0.60 ($\approx 0.588$ actual) respectively.  \par
In previous publications we demonstrated the universal scaling behavior of
conformation and transverse fluctuations~\cite{Huang_JCP_2014} and
crossover dynamics~\cite{Huang_EPL_2014}  of an excluded volume (EV)
swollen chain in 2D. 
We showed that the scaled chain
conformation, $\langle R_N^2 \rangle
/2L\ell_p$  and the  transverse fluctuations $\sqrt{ \langle
  l_{\perp}^2 \rangle}/L$ 
obey universal scaling laws in that, when plotted as a function of 
$L/\ell_p$ both  $\langle R_N^2 \rangle/2L\ell_p$ and $\sqrt{ \langle l_{\perp}^2 \rangle} /L $ for all
combinations of $L$ and $\ell_p$ collapse onto the same master curve
(Figs.~\ref{rg-bulk} and \ref{fluc-bulk}). For $L/\ell_p \ll 1$, in the rod limit, we
observe the expected behavior $ \langle R_N^2 \rangle /2L\ell_p \sim L^2/2L\ell_p
\rightarrow \frac{1}{2}L/\ell_p$. However, for $L\gg \ell_p$, we found
the absence of the Gaussian regime and the scaling behavior of a swollen
chain such that $ \langle R_N^2 \rangle /2L\ell_p \sim
L^{2\nu}/(L^{\nu}\ell_p) \sim (L/\ell_p)^{0.5}$. We interpret that in 2D the extreme
dominance of the EV interaction results in a compete absence of the
Gaussian regime and we observe a direct cross-over from the rod limit
to the EV swollen chain. The universality of the result was further
reassured by the observation that the data from lattice Monte Carlo
simulations using the Prune enriched Rosenblauth scheme by H-P Hsu {\em
  et al.} without any fitting parameter collapsed onto the data
obtained from Brownian dynamics (BD) simulations on the Grest-Kremer bead-spring
model~\cite{Hsu_EPL_2011, Hsu_EPL_2010}. We also provide general arguments regarding
the collapse of the transverse fluctuations onto the same master plot
for all values of $L/\ell_p$. \par

In this article we first generalize and establish those results in
three dimensions (3D) and then extend these studies in the presence of
additional particles which interact with themselves as well as with the
polymer chain with a short range repulsive (EV) interaction for several
different volume fractions. We have also studied the size effect of
the EV particles on these scaling relations. The investigation in 3D is
partly motivated from the theoretical results using the scaling theory of
polymers due to Nakanishi who conjectured that in
3D there will be a broad Gaussian regime before the chain conformation
develops characteristics of a swollen chain for $L \gg \ell_p$~\cite{Nakanishi_1987}. 
Using lattice MC methods, Hsu {\em et al.} demonstrated that for a 3D semiflexible chain there is a Gaussian 
regime which eventually becomes dominated by EV effects. We will
demonstrate that, unlike as depicted in~\cite{Nakanishi_1987}, the width of the Gaussian regime is very narrow, although it can, however, be differentiated from that of a 2D
universal master curve [Fig.~\ref{rg-bulk}(a) and Fig.~\ref{rg-bulk}(b)]. \par 
\section{The Model and the method}
\label{Model0}
\begin{figure}[ht!]
\begin{center}
\includegraphics[width=0.45\textwidth]{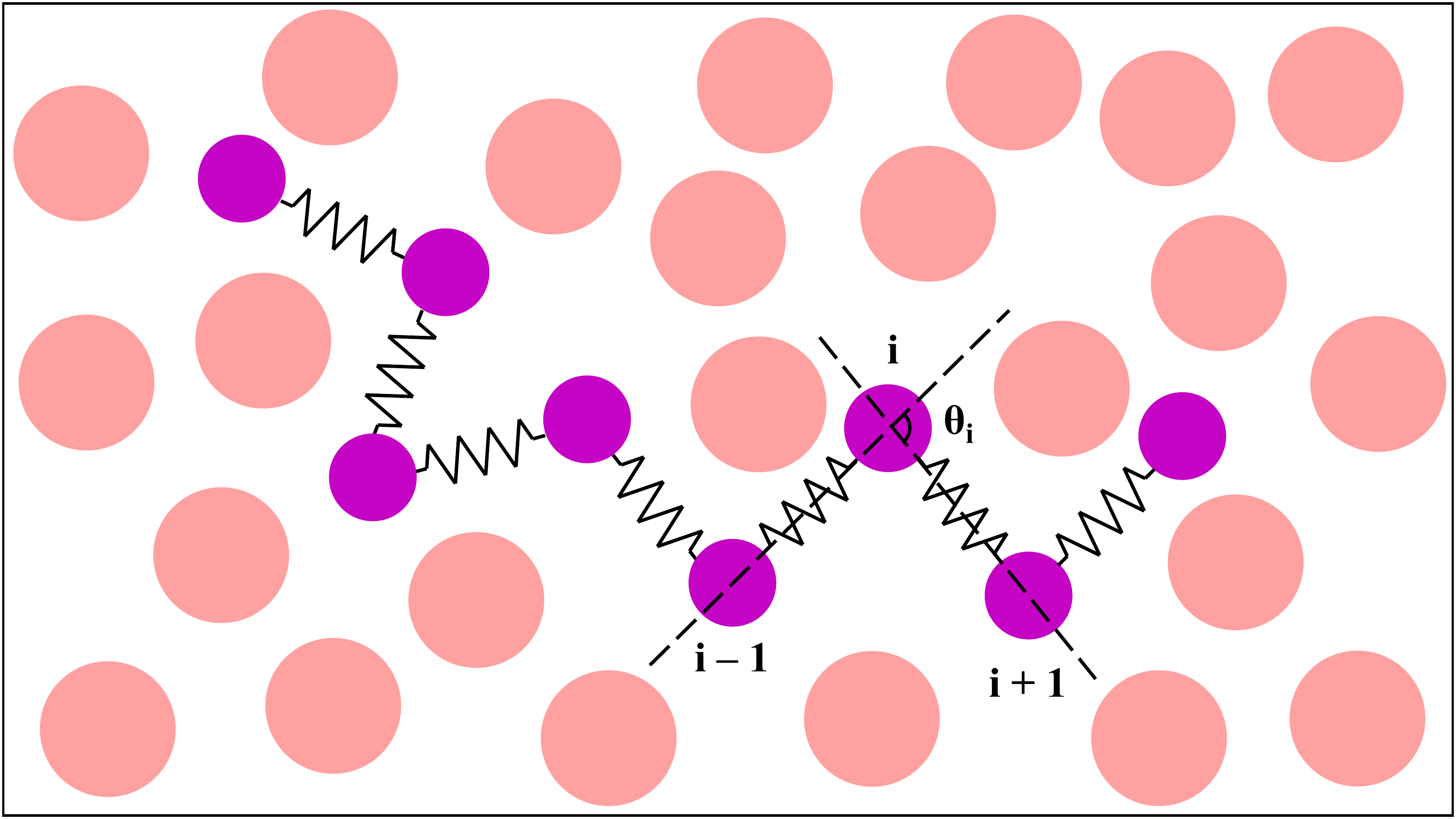}
\end{center}
\caption{\small Schematic shows an eight unit long (N = 8) bead-spring model of a polymer (purple
  beads connected by black springs) in a crowed environment consisting
  of mobile (pink) particles. The
  bond angle for the $i^{th}$ bead is shown as described in Eq.~(\ref{bend}). In the figure
  the diameter of the EV particles $\sigma_{part}=1.5\sigma_{poly}$. }
\label{Model}
\end{figure}

 Our BD scheme is implemented on a Grest-Kremer bead-spring model of a polymer~\cite{Grest} with
 the monomers interacting via an excluded volume (EV), a Finite Extension Nonlinear Elastic (FENE) spring potential, and a
 three-body bond-bending potential that enables variation of the chain persistence
 length $\ell_p$
 [Fig.~\ref{Model} and Eqs.~(\ref{lp_2d}) and (\ref{lp_3d})].

\par
The EV interaction between any two monomers along the chain is given by a short range Lennard-Jones (LJ) potential
\begin{eqnarray}
U_{\mathrm{LJ}}(r)&=&4\epsilon \left[{\left(\frac{\sigma}{r}\right)}^{12}-{\left(\frac{\sigma}
{r}\right)}^6\right]+\epsilon, \;\mathrm{for~~} r\le 2^{1/6}\sigma; \nonumber\\
        &=& 0, \;\mathrm{for~~} r >  2^{1/6}\sigma.
\label{LJ}
\end{eqnarray}
Here, $\sigma=\sigma_{poly}$ is the effective diameter of a monomer on
a polymer chain, and 
$\epsilon$ is the strength of the LJ potential. The connectivity between 
neighboring monomers is modeled as a FENE spring with 
\begin{equation}
U_{\mathrm{FENE}}(r_{ij})=-\frac{1}{2}k_FR_0^2\ln\left(1-r_{ij}^2/R_0^2\right).
\label{FENE}
\end{equation}
Here, $r_{ij}=\left | \vec{r}_i - \vec{r}_j \right|$ is the distance
between the consecutive monomer beads $i$ and $j=i\pm1$ at $\vec{r}_i$
and $\vec{r}_j$, $k_F$ is the spring constant, and $R_0$
is the maximum allowed separation between connected monomers. 
The chain stiffness $\kappa$ is introduced by adding an angle dependent three body interaction term between successive bonds 
as (Fig.~\ref{Model}) 
\begin{equation}
  U_{\mathrm{bend}}(\theta_i) = \kappa\left(1-\cos \theta_i\right).
  \label{bend}
\end{equation}
Here $\theta_i$ is the angle between the bond vectors 
$\vec{b}_{i-1} = \vec{r}_{i}-\vec{r}_{i-1}$ and 
$\vec{b}_{i} = \vec{r}_{i+1}-\vec{r}_{i}$, respectively, as shown in Fig.~\ref{Model}. The strength 
of the interaction is characterized by the bending rigidity $\kappa$
associated with the $i^{th}$ angle $\theta_i$.
For a homopolymer chain the bulk persistence length $\ell_p$ of the chain in
2D and 3D  are given by 
given by~\cite{Landau}
\begin{subequations}
\begin{equation}
  \ell_p/\sigma = 2\kappa/k_BT \quad{\rm (2D)},
    \label{lp_2d}
  \end{equation}
\begin{equation}
\ell_p/\sigma = \kappa/k_BT \quad {\rm (3D)}.
    \label{lp_3d}
  \end{equation}
\end{subequations}
The additional EV particles of diameter $\sigma_{part}$ are introduced 
using the same short-range Lennard-Jones potential with repulsive
cut-off $r_c=2^{1/6}\sigma_{ij}$ as in Eq.~(\ref{LJ}) with $\sigma_{ij}
= \frac{\sigma_i+\sigma_j}{2}$. Here the indices $i$ and $j$ span all
the polymer beads and the additional EV particles with $\sigma_i$ or $\sigma_j$ being either $\sigma_{poly}$ or
$\sigma_{part}$ respectively.\par
We use the Langevin dynamics with the following equations of motion for the i$^{th}$ monomer 
\begin{equation}
m \ddot{\vec{r}}_i = -\nabla (U_\mathrm{LJ} + U_\mathrm{FENE} + U_\mathrm{bend} \\
                                            + U_\mathrm{wall}) -\Gamma \vec{v}_i + \vec{\eta}_i . 
                                          \label{langevin}                                          
\end{equation}

Here $\vec{\eta} _ i (t)$ is a Gaussian white noise with zero mean at temperature $T$, and 
satisfies the fluctuation-dissipation relation in $d=2$ and 3 physical
dimensions.
\begin{equation}
< \, \vec{\eta} _ i (t) 
\cdot \vec{\eta} _ j (t') \, > = 2dk_BT \Gamma \, \delta _{ij} \, \delta (t 
- t ').
\end{equation}
We express length and energy in units of $\sigma$ and $\epsilon$, respectively. 
The parameters for the FENE potential in Eq.~(\ref{FENE}), $k_F$ and 
$R_0$, are set to $k_F = 30 \epsilon/\sigma$ and $R_0 = 1.5\sigma$, respectively. 
The friction coefficient and the temperature are set to 
$\Gamma = 0.7\sqrt{m\epsilon/\sigma^2}$ and $k_BT/\epsilon = 1.2$
respectively. The force is measured in units of $k_BT/\sigma$. The
mass of each bead for both the polymers and the EV particles is chosen to
be the same. 

The numerical integration of Equation~(\ref{langevin}) is implemented
using the algorithm introduced by van Gunsteren and
Berendsen~\cite{Langevin}.   Our previous experiences with BD
simulation suggest that for a time step $\Delta t = 0.01$ these
parameter values produce stable trajectories over  a very long period
of time and do not lead to unphysical crossing of a bond by a
monomer~\cite{Huang_JCP_2014,Huang_JCP_2015}.  The average bond length
stabilizes at $b_l = 0.971 \pm 0.001$ with negligible fluctuation
regardless of the chain size and rigidity~\cite{Huang_JCP_2014}. We
have used a Verlet neighbor list~\cite{Allen} instead of a link-cell
list to expedite the computation. In addition, the simulation runs for
the EV particles were done using LAMMPS~\cite{LAMMPS} with the same
potentials for numerical expediency. We have checked that these runs
yield the same results.
\section{Results}
We first present results for a single semi-flexible chain and the
universal scaling properties in 2D and 3D in Sections A, B, and C. In 
Sections D and E we present the results for the effect of the additional
EV particles.
\subsection{Persistence length and end-to-end distance}
First, we show the results for the universal properties of 
a 3D semiflexible chain. For comparison we have also shown the 2D
results published earlier but with new, added data points~\cite{Huang_JCP_2014}.
\begin{figure}[htb!]
\begin{center}
\includegraphics[width=0.45\textwidth]{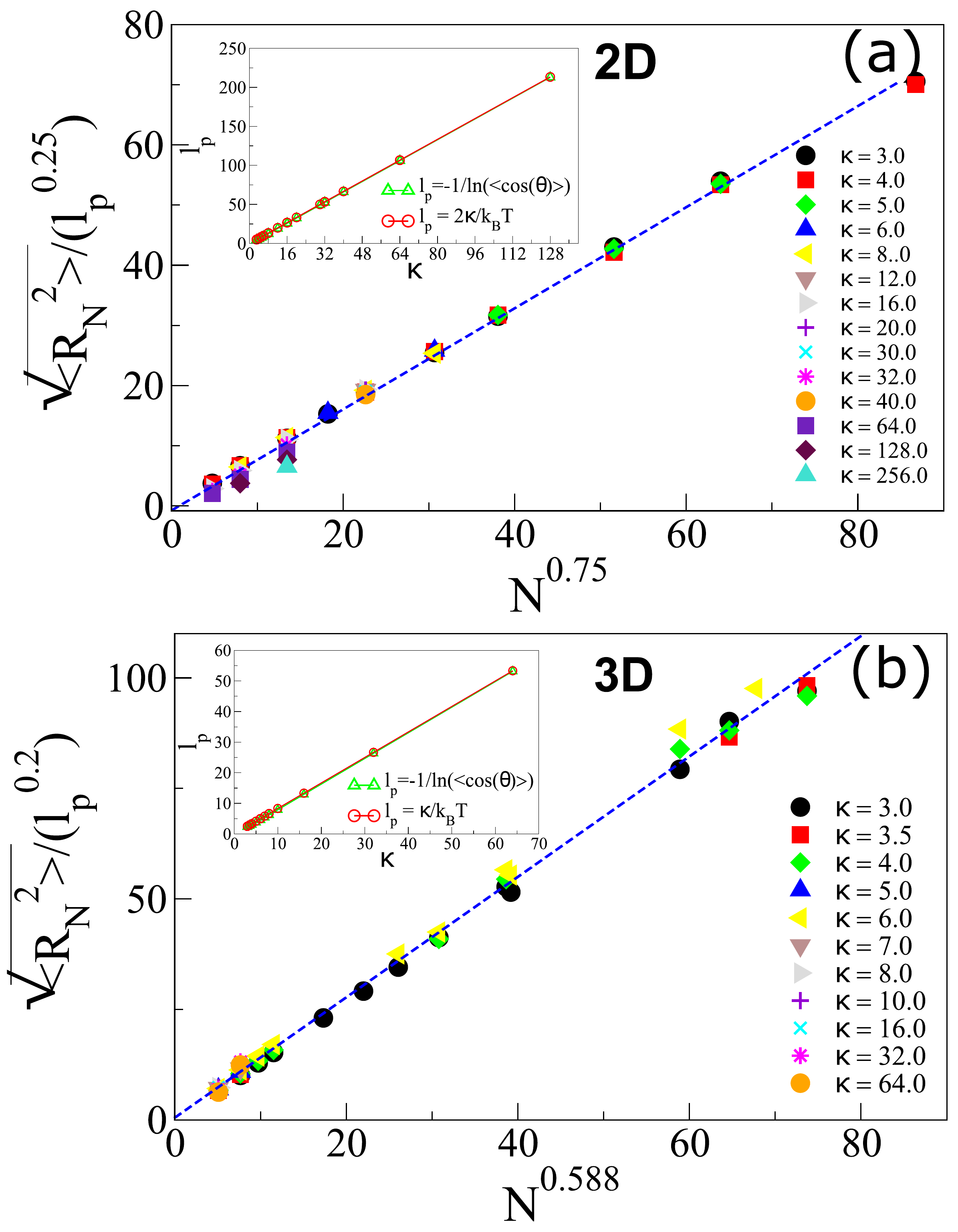}
\end{center}
\caption{\small The scaled end-to-end distance $\sqrt{\langle R_N^2
    \rangle}/l_p^{(1/d+2)}$ as a function of $N^{\nu}$ in (a) 2D and
  (b) 3D. The dashed line in each figure is a straight line fit through
  the points.  The inset in each figures shows the verification of the standard
  definition of persistence length (a) $\ell_p=2\kappa /k_BT$ in 2D [Eq.~(\ref{lp_2d})] 
  and (b) $\ell_p=\kappa /k_BT $ in 3D [Eq.~(\ref{lp_3d})] in presence
  of the EV interaction.}
\label{lp_all}
\end{figure}
\begin{figure}[htb]
\begin{center}
  \includegraphics[width=0.45\textwidth]{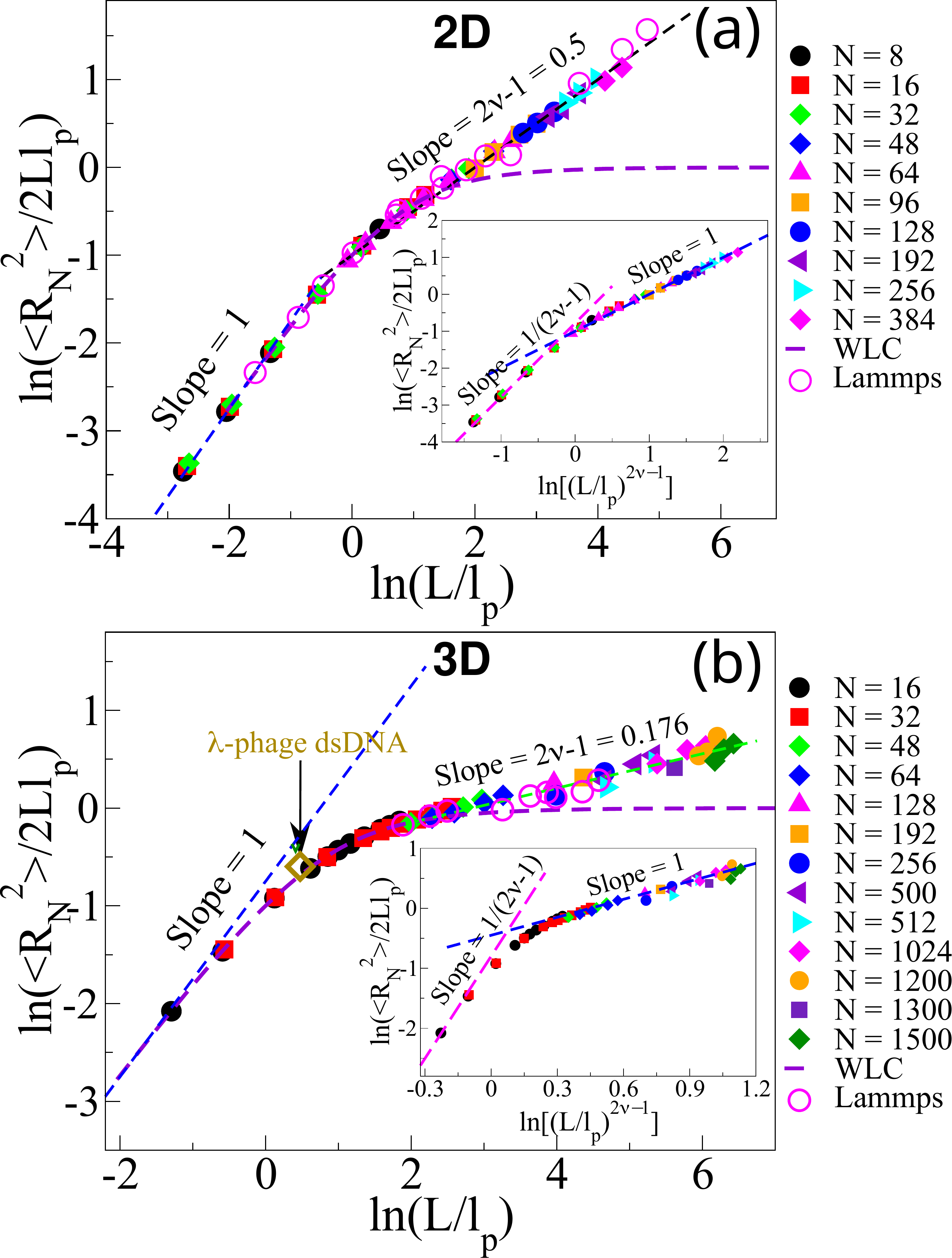}
\end{center}
\vskip -0.25truecm
\caption{\small Log-log plot of the scaled end-to-end distances,
  $\langle R_N^2 \rangle/2Ll_p$, as a function of $L/l_p$ for (a) 2D chains and (b) 3D chains for a variety
  of combinations of $L$ and $\ell_p$. The dashed purple line in each figure shows the
  behavior of the WLC model [Eq.~(\ref{WLC})]. The insets in each
  figure are the plots as a function of
  $(L/\ell_p)^{2\nu-1}$ those show the unit slopes for the
SAW regime and clearly brings out the sharpness of the crossover in 2D and
rounded narrow Gaussian regime in 3D. The symbol
{\textcolor{brown}{\Large{$\diamond$}}} refers to the experimental 
value for $\lambda$-phage
dsDNA from Table I.}
\label{rg-bulk}
\end{figure}
This is required to 
compare with the 2D results in the presence of the EV particles.
A large number of combinations of chain lengths, $N=16 - 512$ for 2D systems and $N=16-1500$ for 3D systems, were chosen. A larger chain
length for the 3D system was necessary to study the crossover from a
Gaussian regime as discussed later. Before we show the scaling results,
we would like to mention that the expression of the persistence length
in Eq.~(\ref{lp_2d}) and (\ref{lp_3d}) is derived
for a WLC~\cite{Landau} but used in Eq.~(\ref{Rn_EV}) to describe a
swollen chain. The validity of the Eqs.~(\ref{lp_2d}) and
~(\ref{lp_3d}) and of Eq.~(\ref{Rn_EV}) for a swollen chain in 2D and
3D are shown in Fig.~\ref{lp_all}. We rationalize the apparent
contradiction by
arguing that the chain persistence length is a local property of the
chain calculated from the immediate three body interaction from a
given bead and thus inclusion of the EV effect does not affect this
result. 
\subsection{Universal aspects of chain conformation}
Now we show the universal aspects of a swollen
chain. Figs.~\ref{rg-bulk}(a)-(b) show the universal scaling and crossover
plots in 2D and 3D respectively. First, we discuss the data collapse of
the root-mean-square (RMS) end-to-end distance $\langle R_N^2\rangle$. The choice of the dimensionless $y$-axis $\langle R_N^2
\rangle/2Ll_p$ in Figs.~\ref{rg-bulk}(a) and \ref{rg-bulk}(b) is guided by noting that in the limit $ L
\gg \ell_p$ Eq.~(\ref{WLC}) results in $\langle R_N^2 \rangle
\rightarrow 2Ll_p$, the Gaussian limit of the WLC. Thus in the absence of
the EV interaction, the quantity $\langle R_N^2 \rangle/2Ll_p \rightarrow 1$ and would exhibit a zero slope [dashed purple line in
Fig.~\ref{rg-bulk}(a) and (b)]. However, for $ L
\gg  \ell_p$, eventually, the EV effect will become important, and from 
Eq.~(\ref{Rn_EV}) it is easy to check that $\langle R_N^2 \rangle \sim
\left (L/\ell_p\right)^{2\nu -1}$. This is clearly the case as
evident from Fig.~\ref{rg-bulk}(a)-(b) for 2D and 3D
respectively. But we note that for 2D there is no Gaussian
regime. For the 3D case, the Gaussian regime is very short. This trend has
also been reported in the MC simulation~\cite{Hsu_EPL_2010} using a
completely different method.
\begin{figure}[ht!]
\begin{center}
  \includegraphics[width=0.48\textwidth]{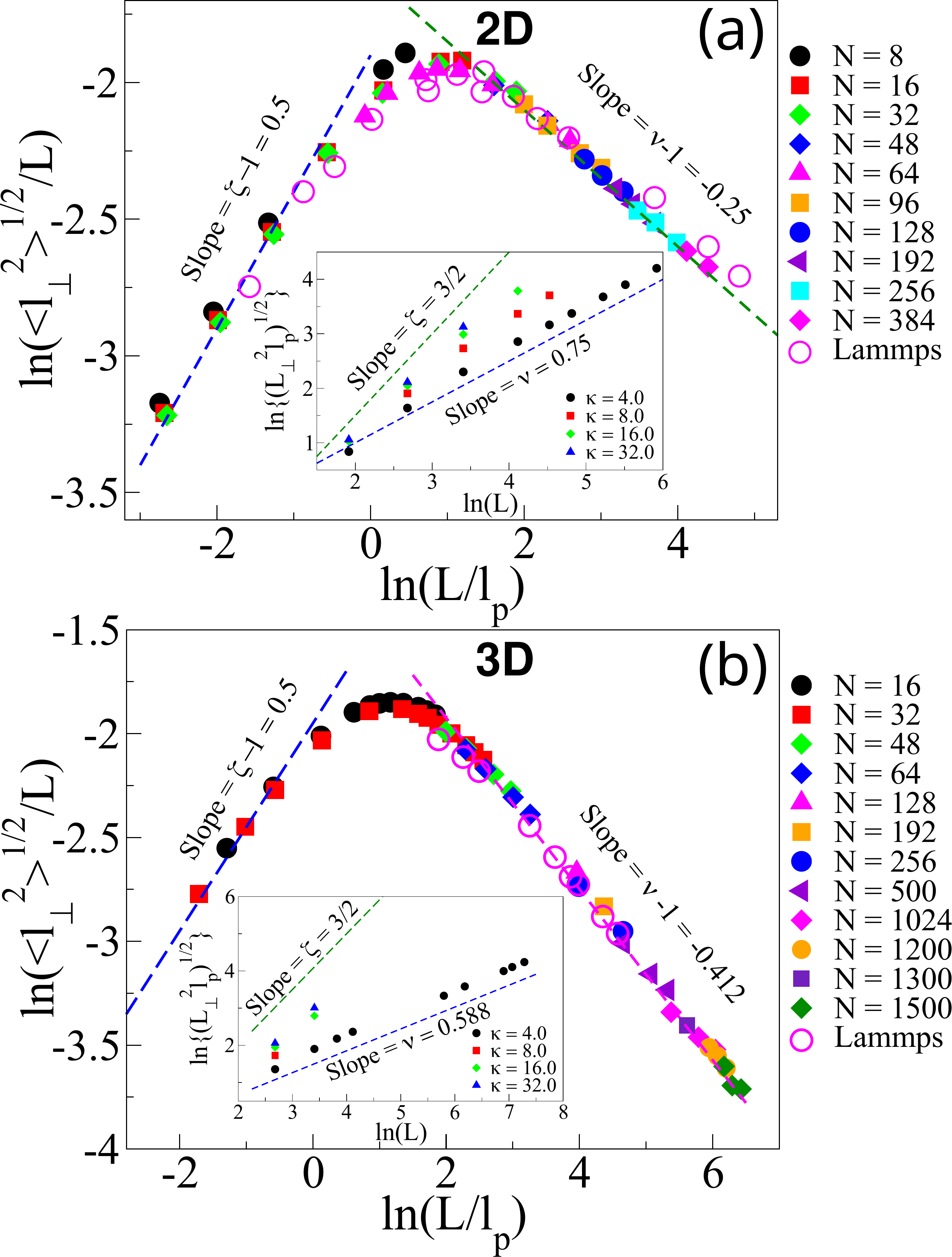}
\end{center}
\vskip -0.5truecm
\caption{\small Log-log plot of the scaled transverse fluctuation
  $\sqrt{\langle l_{\perp}^2 \rangle}/L$, as a function of $L/l_p$ for
  (a) 2D and (b) 3D chains for a variety of combinations
  of $L$ and $\ell_p$. The inset in each case shows the log-log plot of $\sqrt{L_{\perp}^2 l_p}$ as a function of chain's contour length.}
\label{fluc-bulk}
\end{figure}
\subsection{Universal aspects of transverse fluctuations}
We now discuss the universality in transverse fluctuation.
For each configuration of the polymer chain generated during the simulation,
we choose the unit vector $\hat{R}_N=\vec{R}_N/R_N$ as the 
longitudinal axis to calculate transverse fluctuations as follows:  
\begin{equation}
  \langle l_{\bot}^2\rangle = \frac{1}{N}\sum_{i=1}^N y_i^2,
  \label{t_fluc}
\end{equation}
where $y_i$ is the perpendicular distance of the $i^{th}$ monomer with respect to the instantaneous direction $\hat{R}_N$.
We have repeated this calculation for several chain lengths from extremely stiff chains to fully flexible chains.
In the rod limit $\ell_p \gg L$ it can be shown
that for a WLC chain the transverse fluctuation with respect to the
direction of the end-to-end vector obeys the following equation
\begin{equation}
 \langle l_{\perp}^2 \rangle  \sim L^3/\ell_p\;.
 \label{trans-fluc}
\end{equation}
The transverse fluctuation in this limit of a weakly bending rod is
related to the roughness exponent $\zeta$ 
\begin{equation}
\sqrt{\langle l_{\perp}^2 \rangle}  \sim L^{\zeta},
\label{zeta}
\end{equation}
where  $\zeta =\frac{3}{2}$~\cite{Yamakawa}-\cite{Barabasi}.
 Thus, in the limit $\ell_p \gg L$ for an extremely stiff chain, the
transverse fluctuation is governed by the roughening exponent $\langle
l_{\perp}^2 \rangle \sim L^{1.5}$, independent of the spatial dimension
of the system.
In 
the other limit of a fully flexible chain the transverse
fluctuation depends on the physical dimension and is governed by the
Flory exponent of the given spatial dimension as follows.
\begin{equation}
  \langle l_{\perp}^2 \rangle \sim L^{2\nu}.
\label{nu}
\end{equation}
These limits 
are shown in insets of Figs.~\ref{fluc-bulk}(a)-(b). \par 

We also observe that
all the data in Fig.~\ref{fluc-bulk}(a)-(b) collapses onto the same plot
with the peak fluctuation around $L \approx 3\ell_p$. This can be
understood in the following way. The transverse fluctuations go to
zero in the limit of an extremely stiff chain and begin to grow as the ratio $L/\ell_p$ gets
larger. Please note that in order for the ``transverse'' fluctuation to
remain significant compared to the longitudinal fluctuation the chain has to be stiff enough. As the chain
becomes more flexible the fluctuations start to grow in the longitudinal
direction while weakening in the transverse component. 
We find that
when $\ln(L/\ell_p) \simeq 1$, or $L \approx 3\ell_p$, the transverse
fluctuation becomes maximum.
The limiting slopes for the $\sqrt{ \langle l_{\perp}^2
  \rangle}/L$ for $L/\ell_p\ll  1$
and for $L/\ell_p\gg 1$  follow from Eqs.~(\ref{zeta}) and (\ref{nu}) 
can be written as 
\begin{subequations}
  \begin{equation}
   \lim_{L/\ell_p\rightarrow 0} \sqrt{\langle l_{\perp}^2 \rangle}/L
   \sim \left(L/\ell_p\right)^{\zeta-1},
  \end{equation}
  \begin{equation}
    \lim_{L/\ell_p\rightarrow \infty} \sqrt{\langle l_{\perp}^2
      \rangle}/L \sim \left(L/\ell_p\right)^{1-\nu}
  \end{equation}
\end{subequations}
These asymptotic limits
are clearly seen in  Figs.~\ref{fluc-bulk}(a)-(b) both in 2D and in 3D respectively.
The simulation data fits extremely 
well with these predictions. The scaling relation can be used to extract either the $\ell_p$ or the $L$ if one or the other is
known by simply adjusting the ratio $L/\ell_p$ (like a knob) until the point falls
onto the universal plot.
\begin{figure}[htb!]
\begin{center}
\includegraphics[width=0.42\textwidth]{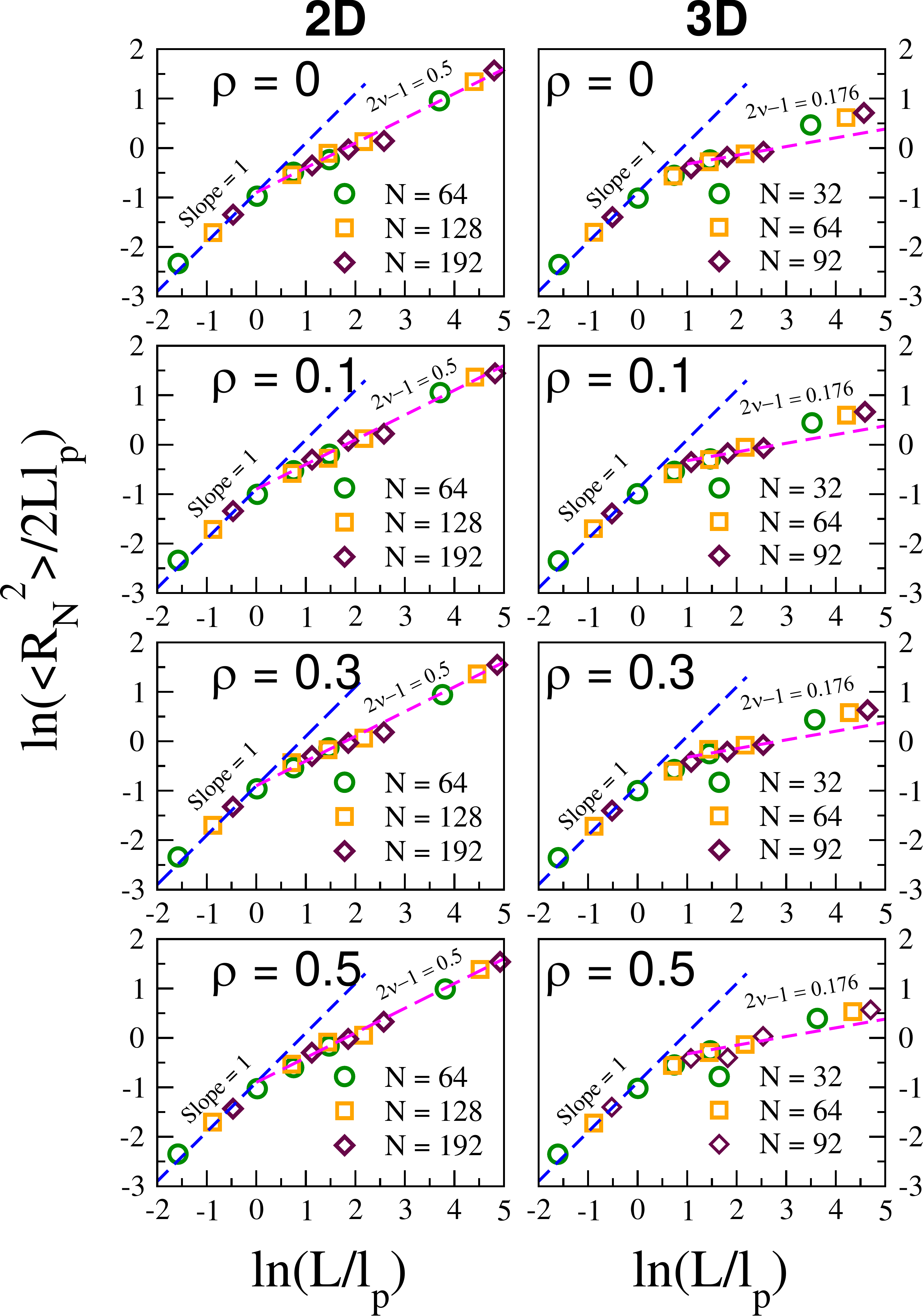}
\end{center}
\caption{\small Log-log plot of the scaled end-to-end distances, 
  $\langle R_N^2 \rangle/2Ll_p$, as a function of $L/l_p$.  The 
  left column (2D) and the right column (3D) 
  correspond to  a variety of combinations of 
  $L$ and $\ell_p$ for several different values of the densities of the crowding particles. The symbols 
  \textcolor{green}{\large $\circ$}, 
  \textcolor{brown}{$\Box$}, 
  and \textcolor{purple}{\large $\diamond$} refer to chain lengths 
  $N=64, 128$ and 192 
  in 2D (left column) and to the chain lengths $N=32, 64$, and 92 
  in 3D (right column) respectively.}
\label{Rg-crowded}
\end{figure}
\subsection{Effect of Crowding}
Having established the universal scaling relations for the
conformations and fluctuations for a single chain, we now study the effect of the EV
particles on these results. The motivation comes from the living world
where biopolymers such as
a double stranded DNA inside a cell encounter crowded environments that affect
their conformation and dynamics and hence their various functionalities. 
In order to check how the universal scaling relations are affected by
the presence of the EV particles, we have studied chains of different lengths ($N=64 - 192$) in the presence of dynamic EV particles of different diameters ($\sigma_{part} = 1.0\sigma, 1.5\sigma$ and $2.0\sigma$) with repulsive
cut-off interaction potentials as a function of the density $\rho=0.1 - 0.5$ (or equivalent
volume fraction $\phi = \frac{1}{4}\rho \pi\sigma_{part}^2$ ) of the
EV particles in both 2D and 3D. We have also varied the chain
persistence length $\ell_p$ such that the ratio $L/\ell_p$ spans a broad
range of values. 
First, we studied the effect of the crowding for EV particles whose
diameters are the same as that of the polymer beads ($\sigma_{part} =
1.0\sigma$, please refer to the Section II and Fig.~\ref{Model}). Then
for two fixed volume fractions ($\phi=0.2356$ and $0.3927$) 
we studied the size effects of EV particles of different diameters
$\sigma_{part} = 1.0\sigma, 1.5\sigma$ and $2.0\sigma$. For this part, we
have carried out BD simulation in 2D only keeping the mass of the beads the same.
\begin{figure}[ht!]
\begin{center}
\includegraphics[width=0.45\textwidth]{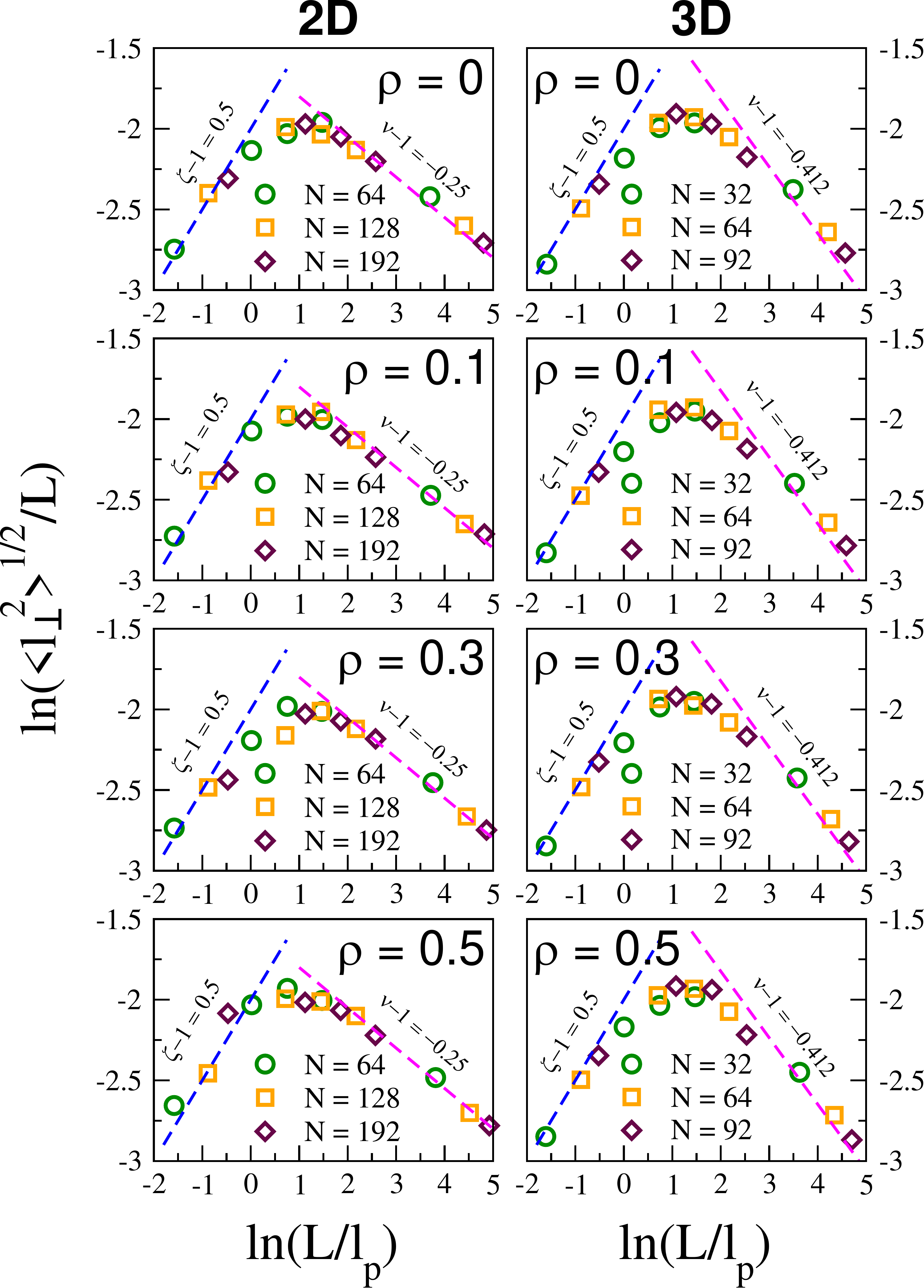}
\end{center}
\caption{\small Log-log plot of the scaled transverse fluctuations, 
  $\sqrt{\langle l_{\perp}^2 \rangle}/L$, as a function of $L/l_p$. The symbols have the same meaning as in Fig.~\ref{Rg-crowded}.}
\label{Fluc-crowded}
\end{figure}
\begin{figure}[htb]
\begin{center}
\includegraphics[width=0.45\textwidth]{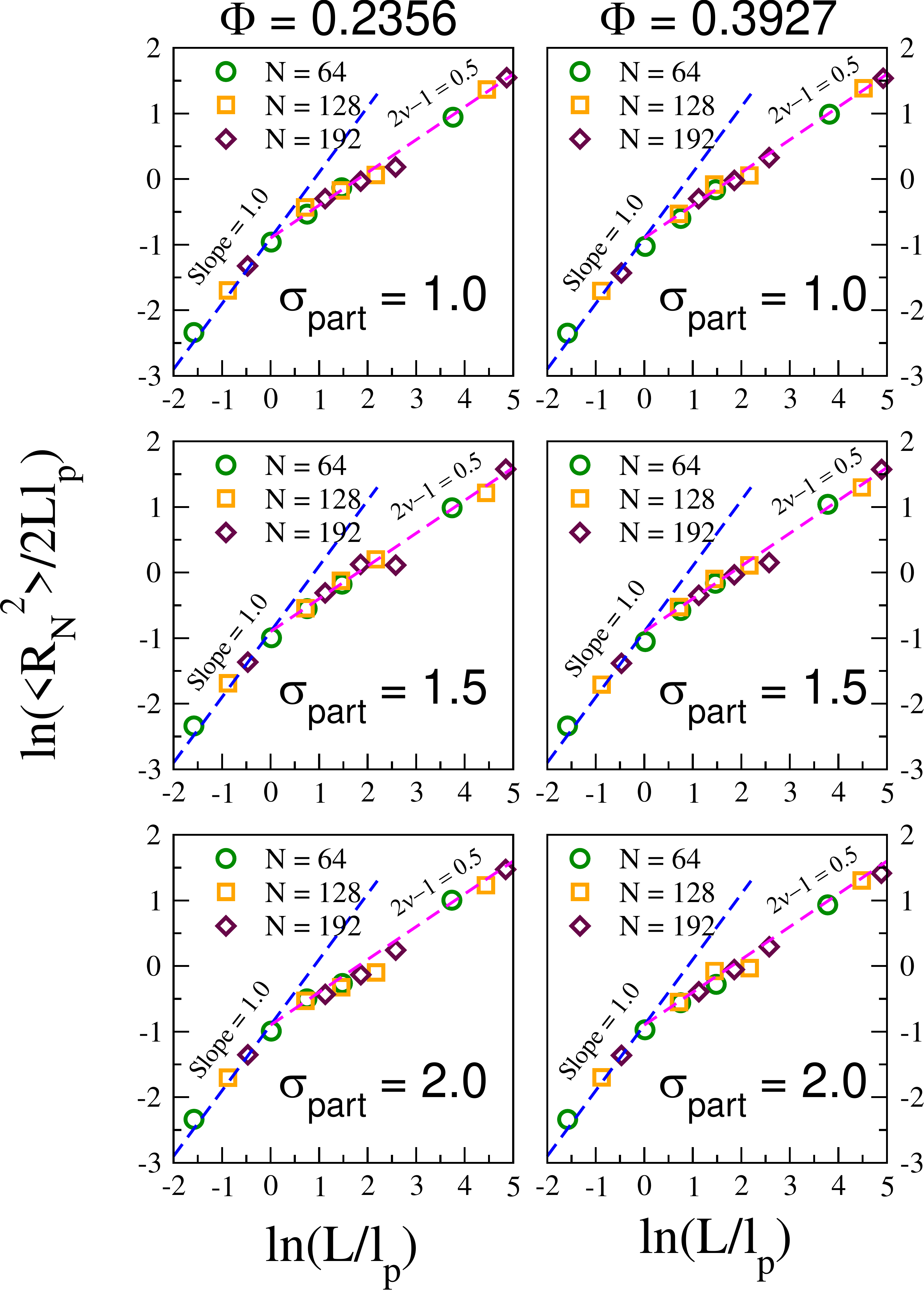}
\end{center}
\caption{\small Log-log plot of the scaled end-to-end distances,
  $\langle R_N^2 \rangle/2Ll_p$ as a function of $L/l_p$ in 2D
  for chain lengths N = 64
  (\textcolor{green}{\large $\circ$}), 
  128 (\textcolor{brown}{$\Box$}), and 192
  (\textcolor{purple}{\large $\diamond$}) for EV particle
  diameters of $\sigma_{part} = \sigma$, 
  $1.5\sigma$, and $2.0\sigma$ respectively. The left and right column
  correspond to two 
  volume fractions $\phi = 0.2356$ and $\phi = 0.3927$ of the
  EV particles.}
\label{rg-rho}
\end{figure}
\subsubsection{Effect of the density of the EV particles of the same
  diameter}
Figs.~\ref{Rg-crowded} and \ref{Fluc-crowded} show the results of the
effect of the additional EV particles of the same diameter on the scaled
end-to-end distance and the transverse fluctuations respectively
both in 2D and in 3D. The
simulations were carried out for particle densities $\rho =
0.1,0.2,0.3, 0.4$, and $0.5$ respectively. Only $\rho = 0.1,0.3$, and
$0.5$ are shown along with $\rho=0.0$ for comparison. For all of the
densities studied, the data indicates
that the EV particles have hardly any effect on the chain
conformations and fluctuations.
\subsubsection{Size effect of the EV particles of different diameters} 
\begin{figure}[htb!]
\begin{center}
\includegraphics[width=0.45\textwidth]{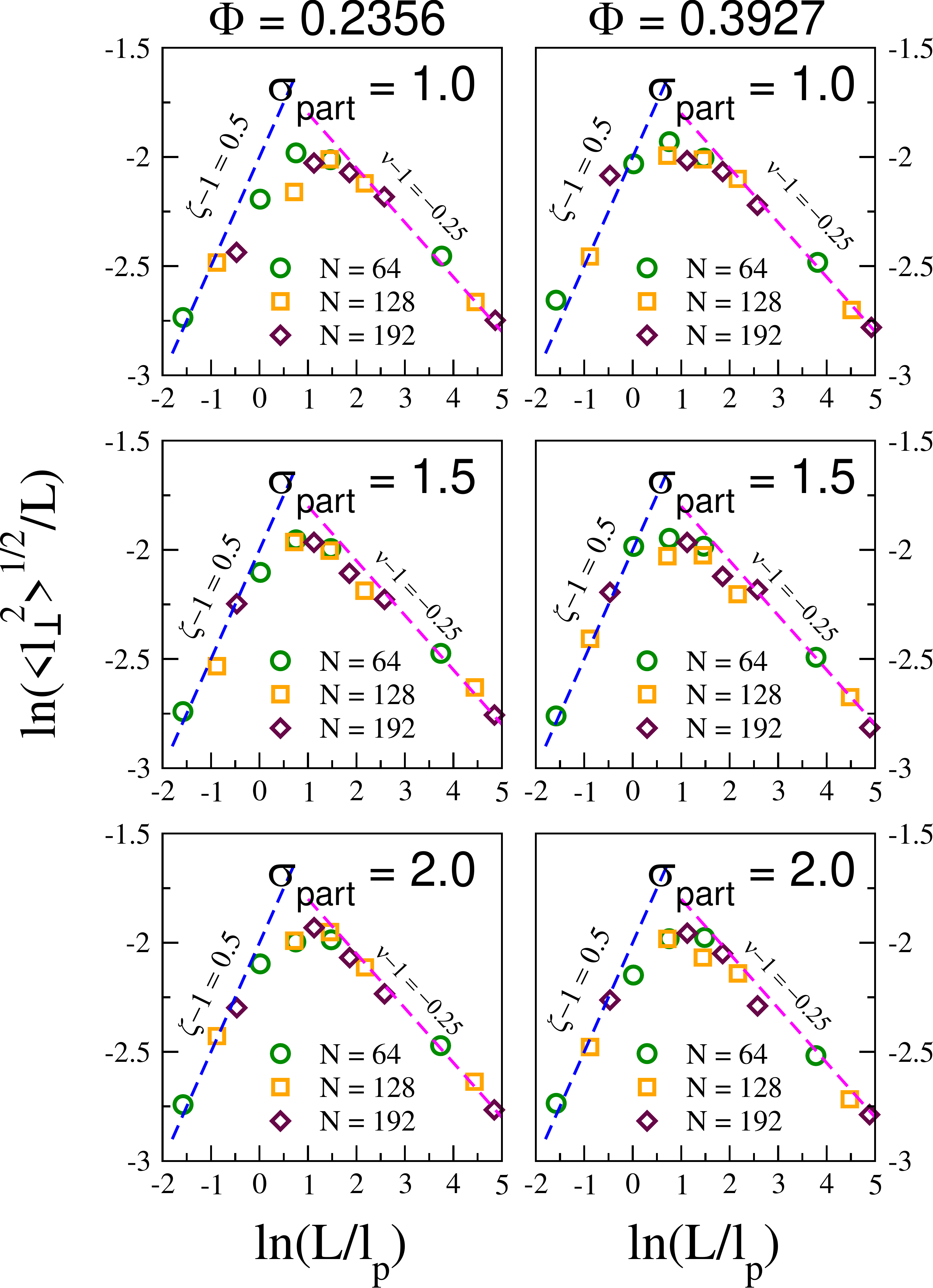}
\end{center}
\caption{\small Log-log plot of the scaled transverse fluctuations,
  $\sqrt{\langle L_{\perp}^2 \rangle}/L$, as a function of the scaled
  lengths, $L/l_p$. The symbols have the same meaning as in Fig.~\ref{rg-rho}.}
\label{fluc-rho}
\end{figure}

In order to investigate how the diameters of the EV particles affect
the results, additional simulations were carried out in 2D for EV particles
with diameters $\sigma_{part} = 1.5\sigma$ and $2.0\sigma$ respectively. We
studied two volume fractions of $\phi = 0.2356$ and $\phi =
0.3927$.
For these simulations, the mass of each EV particle remained the
same. For a volume fraction of $\phi = 0.2356$,
the densities for $\sigma_{part} = 1.0, 1.5$, and 2.0 were $\rho =
0.3, 0.133$, and 0.075 respectively.
For a volume fraction of $\phi = 0.3927$, the densities for particles
with $\sigma_{part} = 1.0, 1.5$, and 2.0 were $\rho = 0.5, 0.22$, and 0.125 respectively.
These results are shown in Figs.~\ref{rg-rho} and
\ref{fluc-rho}. We find that for both the EV particle volume fraction
of $\phi = 0.2356$ and $\phi = 0.3927$, regardless of the diameter
$\sigma_{part}$ of the included EV particles, the data points for each
chain tend to collapse onto the same curve. These results indicate
that, for fixed EV particle volume fractions, the size of the
additional EV particles that are introduced do not appear to
invalidate the scaling relationships up to the maximum diameter of the
EV particles that was tested.\par
We have further investigated the physical origin for the effect of the
EV particles on the scaling laws. 
We have analyzed the simulation data for RMS transverse fluctuation
per unit length  $\langle \sqrt{l_{perp}^2}/L \rangle $ and noticed
that this quantity ($\approx 0.1$) is at least an order of magnitude
less than the average separation among the particles
$\sqrt[3]{L^3/N}=(1.26-1.7)$ for EV particle density $\rho =(0.5–0.2)$. Thus, the presence of the particles on average hardly affects the conformations of the chain. This explains the robustness of the result.  
\subsection{Comparison with experiments}
Biopolymers have a wide range of flexibility. A single stranded DNA
(ssDNA) is more flexible than a double stranded DNA (dsDNA). Actins
and Microtubules are much more rigid. We have gathered 
experimental values of $L/\ell_p$~(\cite{ssDNA}-\cite{Microtubules})
in Table I and determined how they
will be described (rod, Gaussian or a swollen chain) with reference to
the universal scaling plot of Fig.~\ref{rg-bulk}. We notice that a
microtubule and other synthetic stiff polymers are characterized by
rods, while a 30 $\mu$m long Actin~\cite{Actin} filament with a persistence length of
$\ell_p=16.7\;\mu$m falls in the Gaussian regime, while a
ss-DNA~\cite{ssDNA} is described as a swollen chain. For the
$\lambda$-phage dsDNA ($L=75$ nm $\ell_p=46.6$nm)  we also have an experimental value for the mean-square end-to-end distance,
${\langle R_N^2 \rangle} = 3844$, nm that enables us to obtain the
scaled coordinate
$(L/\ell_p, {\langle R_N^2\rangle} /2L\ell_p)\equiv (1.61,
0.55)$. This coordinate falls
right onto the universal plot ({\color{brown}{\Large{$\diamond$}}}) in
Fig.~\ref{rg-bulk} in the Gaussian regime and serves as a testimonial to
our theory. It is worth
making a comment in this context that most of the biopolymers are
described as WLC. But, a large number of them will behave as swollen
chains. The universal curve of Fig.~\ref{rg-bulk} can be used to
classify them.  Fig.~\ref{fluc-bulk} can then be used to extract the transverse
fluctuations of the chains.
\begin{table}[htb!]
\label{Table1}
\centering  
\begin{tabular}{|c|c|c|c|c|}
\hline  
Polymer & L & $l_p$ & $ln\left( \frac{L}{l_p} \right)$ & \textbf{Regime} \\
\hline  
Microtubules~\cite{Microtubules} & 28.9 $\mu$m & 5.2 mm & -5.193 & Rod \\ 
BCHV-PPV~\cite{Polyvinyl}     & 0.6 nm & 40 nm & -4.199 & Rod \\ 
BEH-PPV~\cite{Polyvinyl}       & 0.6 nm & 11 nm & -2.909 & Rod \\ 
MEH-PPV~\cite{Polyvinyl}      & 0.6 nm & 6 nm & -2.303 & Rod \\ 
Actin~\cite{Actin}      & 30 $\mu$m & 16.7  $\mu$m & 1.796 &
                                                                       Gaussian\\
  $\lambda$-phage dsDNA~\cite{dsDNA} & 75 nm & 46.6 nm & 0.476 & Gaussian \\ 
ssDNA~\cite{ssDNA}        & 120.4 nm & 3.1 nm & 3.13 & SAW \\
             & 2316 nm & 5.2 nm & 6.61 & SAW \\
\hline  
\end{tabular}
\hspace{0.5cm}
\caption{Experimental values of the contour length $L$ and persistence  
  length $\ell_p$ of various semi-flexible bio and synthetic polymers  
  in the bulk and their description using the scaling plot of  
  Fig.~\ref{rg-bulk}(b). For the  $\lambda$-phage dsDNA, 
  $\langle R_N^2 \rangle = 3844$ nm~\cite{dsDNA} and falls onto  
  the Gaussian regime of the master plot of 
  Fig.~\ref{rg-bulk}(b) [the symbol
{\textcolor{brown}{\Large{$\diamond$}}} in Fig.~\ref{rg-bulk}(b)].}
\end{table}
\section{Summary and conclusion}
In conclusion, we have established universal aspects of conformations and
fluctuations of a semi-flexible chain by studying the scaled
end-to-end distance $ \langle R_N^2 \rangle /2L\ell_p$ and the scaled
transverse fluctuation $\sqrt{\langle l_{\perp}^2 \rangle }/L$ as a
function of the scaled contour length $L/\ell_p$ . The purpose of the
choice for the former is that in the limit of a flexible chain it
exhibits the characteristics of a Gaussian chain and thus the effect
of the EV will become immediately observable. The purpose of the choice of
the latter is that the root-mean-square fluctuation per unit length is
what is important. Furthermore, the relative flexibility of the chain
is measured in units of $L/\ell_p$ and should be the 
correct length unit to understand the results. Therefore, when plotted
as a function of 
$L/\ell_p$ both the end-to-end distance and the transverse
fluctuations collapse onto the master plots.
Comparing the plots in 2D and 3D, we conclude that the Gaussian regime,
though present in 3D, is very narrow.
Therefore, most of the long
semiflexible polymers will be characterized by a swollen chain.

The transverse fluctuations, as expected in the rod limit, are independent of the spatial dimensions
and grow as $\sqrt{\langle l_{\perp}^2 \rangle }\sim L^{3/2}$  as
described by the roughening exponent $\zeta = 1.5$ (Fig.~\ref{fluc-bulk}) while in
the limit of a flexible chain the fluctuation is dimension dependent
and grows as $\sqrt{\langle l_{\perp}^2 \rangle }\sim L^{\nu}$,
where $\nu$ is the corresponding Flory exponent in a given
dimension. We extend our previous work and observe that not only for the
asymptotic limits but for all ratios of $L/\ell_p$ both the scaled end-to-end distance and the scaled
fluctuations collapse on universal plots, indicating that the
appropriate length
scale to analyze the data is $L/\ell_p$ which brings out these universal aspects. \par

Moreover, we observe that crowding due to the EV particles of different
volume fractions and of the same and different sizes does not change the universality of
these results. We understand this by noting that the magnitudes of the scaled transverse
fluctuations are much less than the average separation of the EV
particles indicating that on average chain fluctuations and
conformations are hardly affected by the EV particles for the
densities studied here. Thus, we believe these results will be useful to
calibrate and characterize both semiflexible biopolymers and synthetic
polymers with respect to a universal scale. We also would like to
point out that we haven't varied the mass of the spherical
particles as a function of their sizes and have not included the
hydrodynamics effects in this study. Varying the mass of the spherical
particles only is not likely to change the universal scaling laws as
they are polymer specific and are not affected by the presence of the
particles of equal mass. However, HD effects may change these results
which is beyond the scope of these studies. We conclude by stating
that these results can
be used as references to classify properties of intrinsically disordered
proteins (IDP) which remain in an extended state and whose studies have become an
increasingly important and emerging field~\cite{IDP}.

\section*{Acknowledgments}
All calculations have been carried out
under the auspices of UCF's high-performance cluster Stokes. AB thanks
late Prof. Kurt Binder for various discussions on this problem.


\end{document}